\documentclass[11pt,twoside]{article}

\usepackage{asp2004}
\usepackage{epsf}
\usepackage{psfig}
\usepackage{lscape}
\usepackage{graphicx}

\begin{document}
\title{Single massive stars at the critical rotational velocity: possible links with
Be and B[e] stars}   
\author{Georges Meynet and Andr\'e Maeder}   
\affil{Geneva Observatory}    

\begin{abstract} 
Using single star models including the effects of shellular rotation with and without
magnetic fields, we show that 
massive stars at solar metallicity with initial masses lower than about 20-25 M$_\odot$ 
and with an initial rotation above $\sim 350$ km s$^{-1}$
likely reach the critical velocity during their Main-Sequence phase. 
This results from the efficient outwards transport of angular momentum by the meridional circulation. This could be a scenario for explaining the Be stars.
After the Main-Sequence phase, single star in this mass range can again reach the critical limit when they
are on a blue loop after
a red supergiant phase \citep{HL98}. This might be a scenario for the formation of B[e] stars, however
as discussed by Langer \& Heger~(1998), this scenario would predict a short B[e] phase (only some 10$^4$ years) with
correspondingly small amounts of mass lost.  

\end{abstract}


\section{Link between Be, B[e] supergiants stars and rotation}

A common feature of Be and B[e] supergiants is the non-sphericity of their circumstellar envelopes
(see e.g. the review by Zickgraf~2000). More precisely, 
in both cases, disks are supposed to be present, likely disk of outflowing material. 
How do these disk form~? How long are their lifetimes~? Are they intermittent~? 
Are they Keplerian~? Many of these questions have been discussed in 
this conference and are still subject of lively debate. A point however which seems well accepted is the fact that
the origin of an axisymmetric wind structure such as a disk might be connected to the fast
rotation of the star \citep{Pe00}. 
If correct, this connection between fast rotation and the Be and B[e] 
phenomena leads to the question, when such fast rotation can be 
encountered in the course of the evolution of massive single stars ? 
In this paper, we give some elements of answer to that question based on models accounting
for the effects of shellular rotation \citep{Za92}. 


\section{The critical velocity}

The critical angular velocity corresponds to the angular velocity at the equator 
of the star such that the centrifugal force exactly balances the gravity.
The critical angular velocity $\Omega_{\rm crit,1}$ in the 
frame of the Roche model for computing the gravity
due to the deformed star, is given by 
\begin{equation}
\Omega_{\rm crit,1}=\left({2 \over 3}\right)^{3 \over 2}\left({GM \over R^3_{\rm pc}}\right)^{1 \over 2},
\label{eqn1}
\end{equation}
where $R_{\rm pc}$ is the polar radius when the surface rotates with the critical velocity.

Looking at Eq.~(\ref{eqn1}), one can be surprised that the stellar luminosity does not appear.
Indeed, we could expect that, in addition to the centrifugal acceleration, the radiative acceleration 
would help in balancing the gravity. When the stellar luminosity is sufficiently far from the
Eddington limit (see below for a more precise statement), it has been shown by Glatzel~(1998) and
Maeder \& Meynet~(2000) that radiative acceleration
does not play any role. Physically, this comes from the fact that when the star reaches the critical limit at the equator, the
effective gravity (gravity decreased by the effect of the centrifugal acceleration) 
becomes zero there and the radiative flux, responsible for the radiative acceleration, tends  also toward
zero due to the von Zeipel theorem (von Zeipel~1924; Maeder~1999).

\begin{figure}[!t]
\plottwo{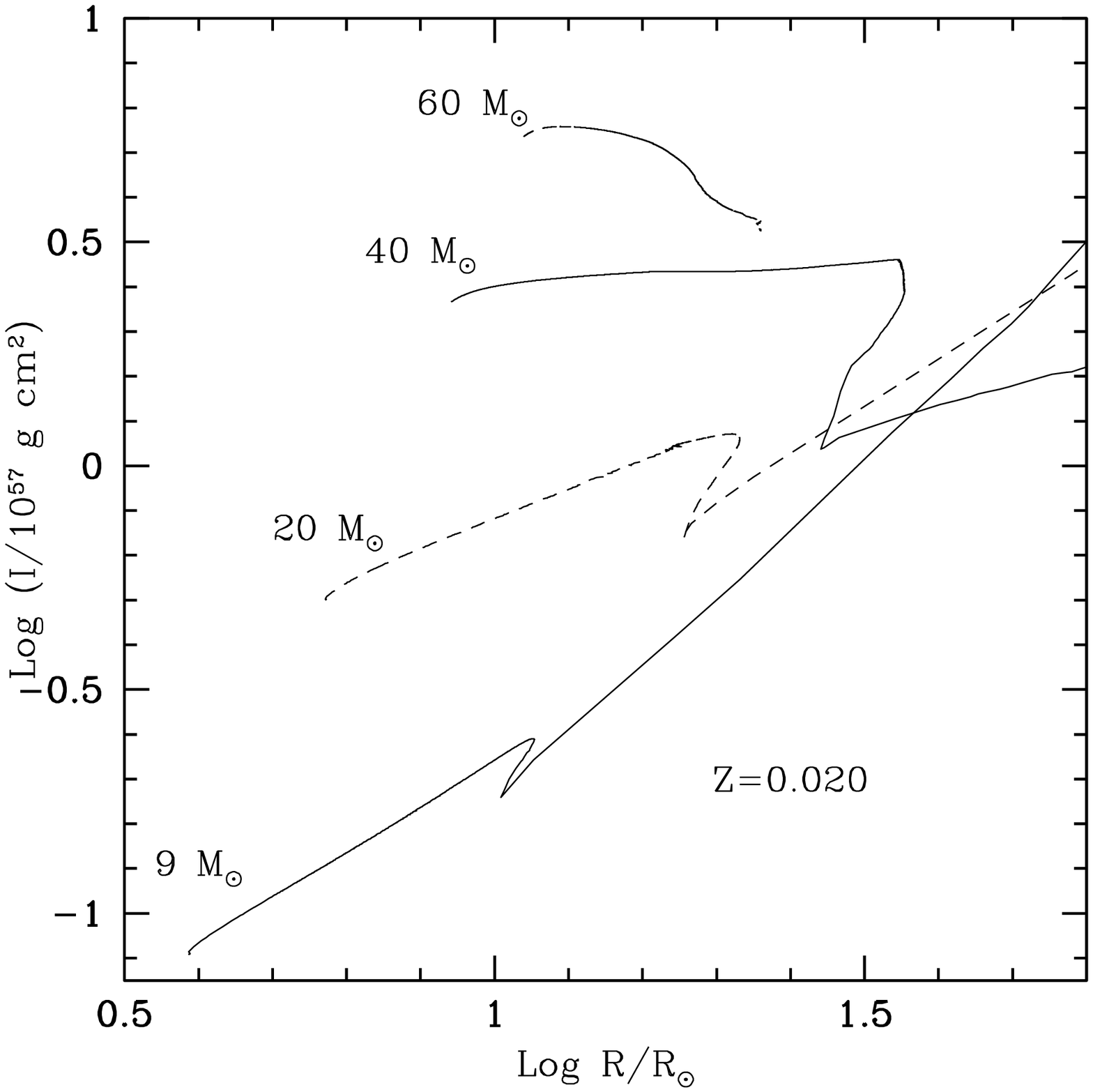}{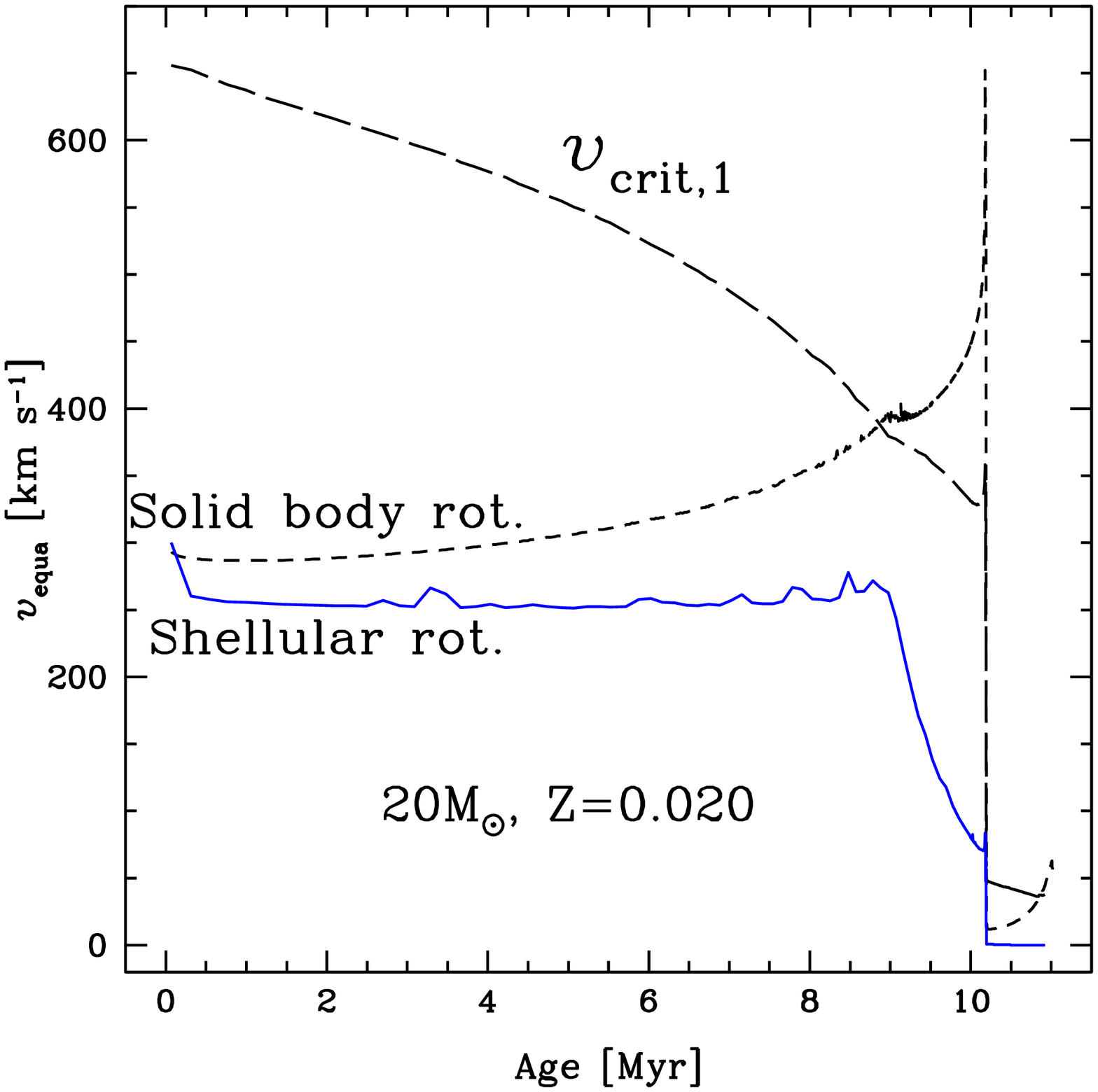}
\caption{{\it Left :} Evolution of the total momentum of inertia of different stellar models at solar
metallicity. The hook along the tracks corresponds to the end of the Main Sequence phase.
Only the beginning of the track is shown for the 60 M$_\odot$ stellar model.{\it Right :} The continuous line shows
Evolution of the surface velocity of a 20 M$_\odot$ at solar metallicity with $\upsilon_{\rm ini}=300$
km s$^{-1}$ according to the models computed by Meynet \& Maeder~(2003). The long--dashed line
corresponds to the critical velocity given by Eq.~\ref{eqn1}. The short--dashed line
shows the surface velocity the star would have, if solid body rotation is assumed.}
\label{momi}
\end{figure}

In contrast, when the stellar luminosity approaches the Eddington limit, the radiative acceleration becomes
a dominant effect. Why such a difference between the case far from 
the Eddington limit and the case near the Eddington limit ? 
One could indeed argue that even near the Eddington limit, 
when the critical limit is approached, the radiative 
flux becomes zero at the equator. This is correct, 
but another mechanism comes into play here: the fact 
that the Eddington limit is modified when the star is rotating. Let us first recall that the classical Eddington luminosity
is given by the expression
\begin{equation}
L_{\rm Edd}=4\pi c G M/\kappa,
\label{eqn2}
\end{equation}
where $\kappa$ is the total opacity, 
$L$ the luminosity, $M$ the mass of the star and the other symbols have their usual meanings.
Now, when the star is rotating, 
two important differences appear: first the Eddington limit varies 
as a function of the colatitude $\theta$, second, it is decreased when the rotational velocity increases. The
Eddington limit modified by rotation is given by (Glatzel 1998; Maeder \& Meynet 2000)
\begin{equation}
L_{\rm Edd}(\Omega)=4\pi c G M\left(1-{\Omega^2 \over 2\pi G \overline{\rho}}\right)/\kappa(\theta),
\label{eqn3}
\end{equation}
where $\overline{\rho}=M/V(\omega)$ is the average density of the star, $V(\omega)$ being the stellar volume
when the star is rotating and $\omega$ is the ratio of the angular velocity to the critical angular velocity given in Eq.~(\ref{eqn1}).

\begin{figure}[!t]
\plottwo{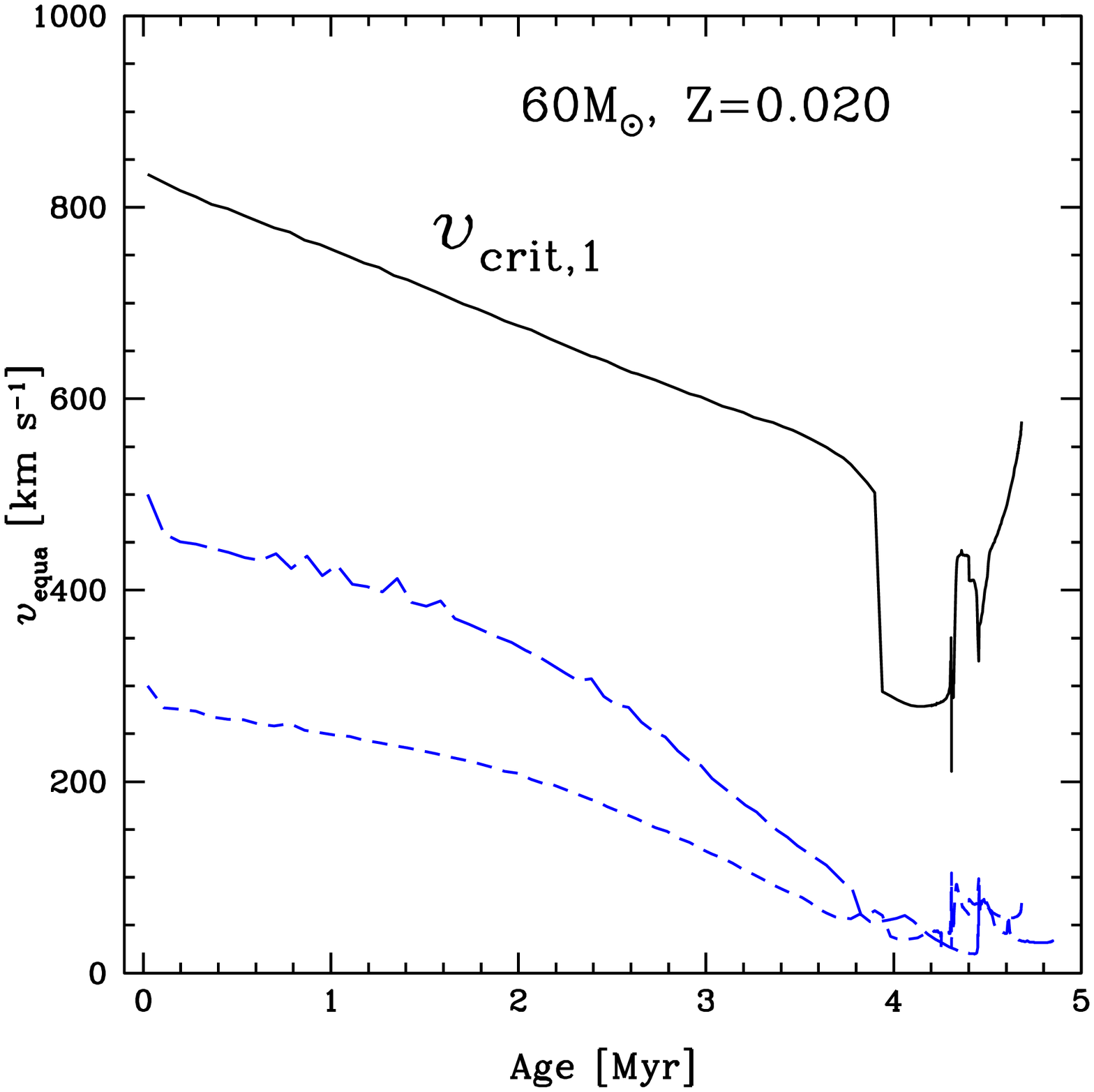}{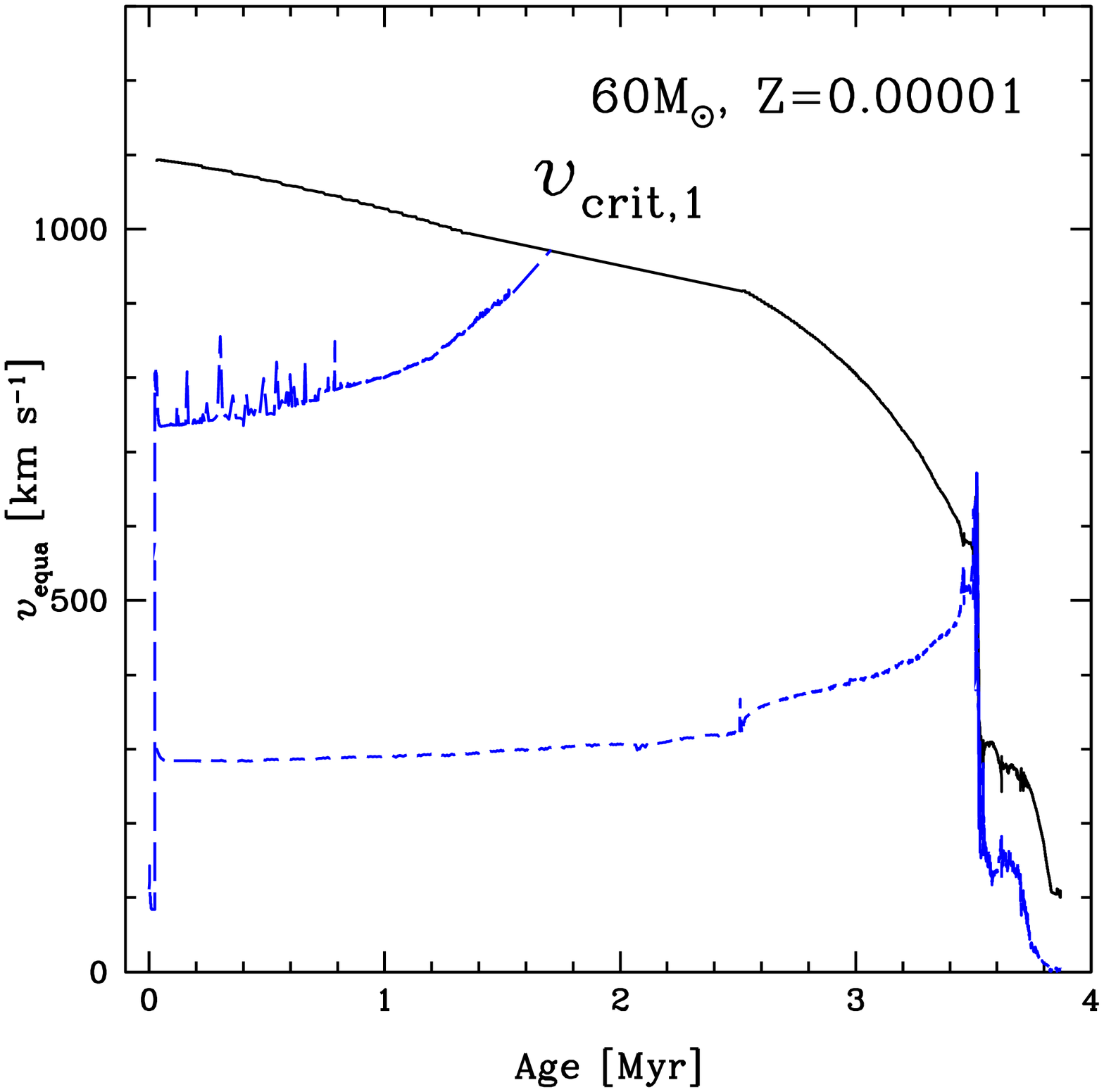}
\caption{{\it Left :} Evolution of the surface equatorial velocities at the surface of 60 M$_\odot$ models at solar metallicity with $\upsilon_{\rm ini}=$ 300 (short-dashed line) and 500 km s$^{-1}$ (long-dashed line). The continuous line shows the evolution of the equatorial critical velocity given by Eq.~(\ref{eqn1}).
{\it Right :} Same as the left part of the figure, for 60 M$_\odot$ at $Z=10^{-5}$ with $\upsilon_{\rm ini}=$ 300 (short-dashed line) and 800 km s$^{-1}$ (long-dashed line).}
\label{vz}
\end{figure}

Thus, when the star is near the Eddington limit, rotation may sufficiently decrease $L_{\rm Edd}$ to 
make it equal to the actual luminosity of the star. In that case, we say the star has reached the
$\Omega\Gamma$-limit and strong mass loss ensues.

Now at which velocity does this occur ? 
To obtain it, one has to find the value of $\Omega$ such that
$L=L_{\rm Edd}(\Omega)$, where $L_{\rm Edd}(\Omega)$ is given by Eq.~(\ref{eqn3}). This is equivalent to find $\Omega$ such that
\begin{equation}
\Gamma_{\rm max}\equiv{\kappa_{\rm max} L \over 4\pi c G M}=1-{\Omega^2 \over 2\pi G \overline{\rho}},
\label{eqn4}
\end{equation}
where we have added the subscript ``max'' to indicate that the critical limit will be reached first
at the position on the surface where the opacity is maximum.
If the value of $\Omega$ satisfying this equality
is higher than $\Omega_{\rm crit,1}$
(from now on called the classical limit),
then only the classical limit is relevant since it will be reached first.
To see if Eq.~(\ref{eqn4}) can be fulfilled with $\Omega < \Omega_{\rm crit,1}$, let
us use the definition of $\overline{\rho}$ and of $\omega$ to write
\begin{equation}
{\Omega^2 \over 2\pi G \overline{\rho}}={16 \over 81}\omega^2 V'(\omega),
\label{eqn5}
\end{equation}
with $V'(\omega)={V(\omega) \over {4 \over 3} \pi R^3_{\rm pc}}$.
From the Roche model, it can be shown that the
quantity ${16 \over 81}\omega^2 V'(\omega)$ increases from 0 to 0.361 when $\omega$ varies from
0 to 1 (see Maeder \& Meynet~2000). It means that if $\Gamma_{\rm max}$ is strictly inferior to
1 - 0.361 = 0.639, Eq.~(\ref{eqn4}) cannot be fulfilled. If $\Gamma_{\rm max}=0.639$, Eq.~(\ref{eqn4})
can be fulfilled with 
$\Omega = \Omega_{\rm crit,1}$ and if 
$\Gamma_{\rm max}$ is superior to 0.639, then values of
$\Omega < \Omega_{\rm crit,1}$ can satisfy Eq.~(\ref{eqn4}). A new expression
for the critical velocity, valid when the star is sufficiently near the Eddington limit, can be derived. It is given by
\begin{equation}
\Omega_{\rm crit,2}=\left({9 \over 4}\right)\Omega_{\rm crit,1} \sqrt{{1-\Gamma_{\rm max}\over V'(\omega)}}
\label{eqn6}
\end{equation}
where $R_{\rm e}(\omega)$ is the equatorial radius for a given value
of the rotation parameter $\omega$. 
These expressions for the critical velocities are different from the expression
\begin{equation}
\Omega_{\rm crit}=\Omega_{\rm crit,1} (1-\Gamma)
\label{eqn7}
\end{equation}
used by some authors. Expression~(\ref{eqn7}) is correct only if the surface is uniformly bright, which is not the case when the star is rotating fast.

\section{Evolution of the surface velocity: two simple extreme cases}   

The evolution of the rotational velocity at the surface of stars depends mainly on
three physical processes:
\begin{itemize}
\item The efficiency of the angular momentum transport mechanisms in the interior,
\item The movement of expansion/contraction of the surface,
\item The mass loss.
\end{itemize}

An extreme case of internal angular momentum transport is the one which imposes solid
body rotation at each time in the course of the evolution of the star. A strong coupling is then realised
between the contracting core and the expanding envelope. In that case, 
the
angular velocity $\Omega$ is given by the ratio of the total angular momentum $J$ and the total
momentum of inertia of the star $I$. Fig. \ref{momi} shows the variation of $I$ as a function
of the growing stellar radius for a few stellar models at solar metallicity.
In the case of the 9 M$_\odot$ model, $I$ varies as $R^\alpha$ with $\alpha\sim 1$.
In the case of the 60 M$_\odot$ model, $\alpha$ becomes negative. This results from the strong
mass loss experienced by this star.

For the 9 M$_\odot$, since mass loss by stellar winds remains very modest during the Main Sequence phase, the total
angular momentum is conserved during this phase. As a consequence, $\Omega$ varies as the inverse of $I$ and 
since $I\propto R$, $\Omega\propto 1/R$. 




From the previous section, we saw that when the star is sufficiently far from the Eddington limit,
which is the case for the 9 M$_\odot$ stellar model, then $\Omega_{\rm crit,1} \propto R^{-3/2}$.
Thus the critical angular velocity decreases more rapidly than the surface
angular velocity  when the star expands. Clearly this favors the reaching of the critical limit 
(see also Sackmann and Anand~1970; Langer~1998).

To illustrate this last point, let us consider the rotating track for the 20 M$_\odot$ model (solar metallicity and
initial rotational velocity of 300 km s$^{-1}$)
computed by Meynet \& Maeder~(2003) for obtaining
values of the momentum of inertia during the evolution.
From these values of $I$ and also using the values for the actual total angular momentum, we deduce the surface
velocity that the star would have in case of solid body rotation. We obtain 
the short dashed line in Fig.~\ref{momi}. Although the model is not self consistently computed
(the evolutionary tracks were not computed imposing solid body rotation),
it illustrates the fact that indeed when solid
body rotation is achieved, the star may reach very easily the critical limit
(here represented by the long--dashed line). 

Another extreme case is the case of no transport of angular momentum. Each stellar layer keeps its
own angular momentum. 
The variation of $\Omega$ is then simply governed by the local conservation of the
angular momentum and, at the surface, $\Omega \propto 1/R^2$. 
When the radius increases, the surface angular velocity decreases 
more rapidly than the classical critical angular velocity, thus
the star evolves away from the critical limit.

Reality is likely in between the cases of solid body rotation and of
local conservation of the angular momentum.
Let us now see what
are the predictions of more physical models.

\section{Evolution of the surface velocity in models with shellular rotation}

\begin{figure}[!t]
\plottwo{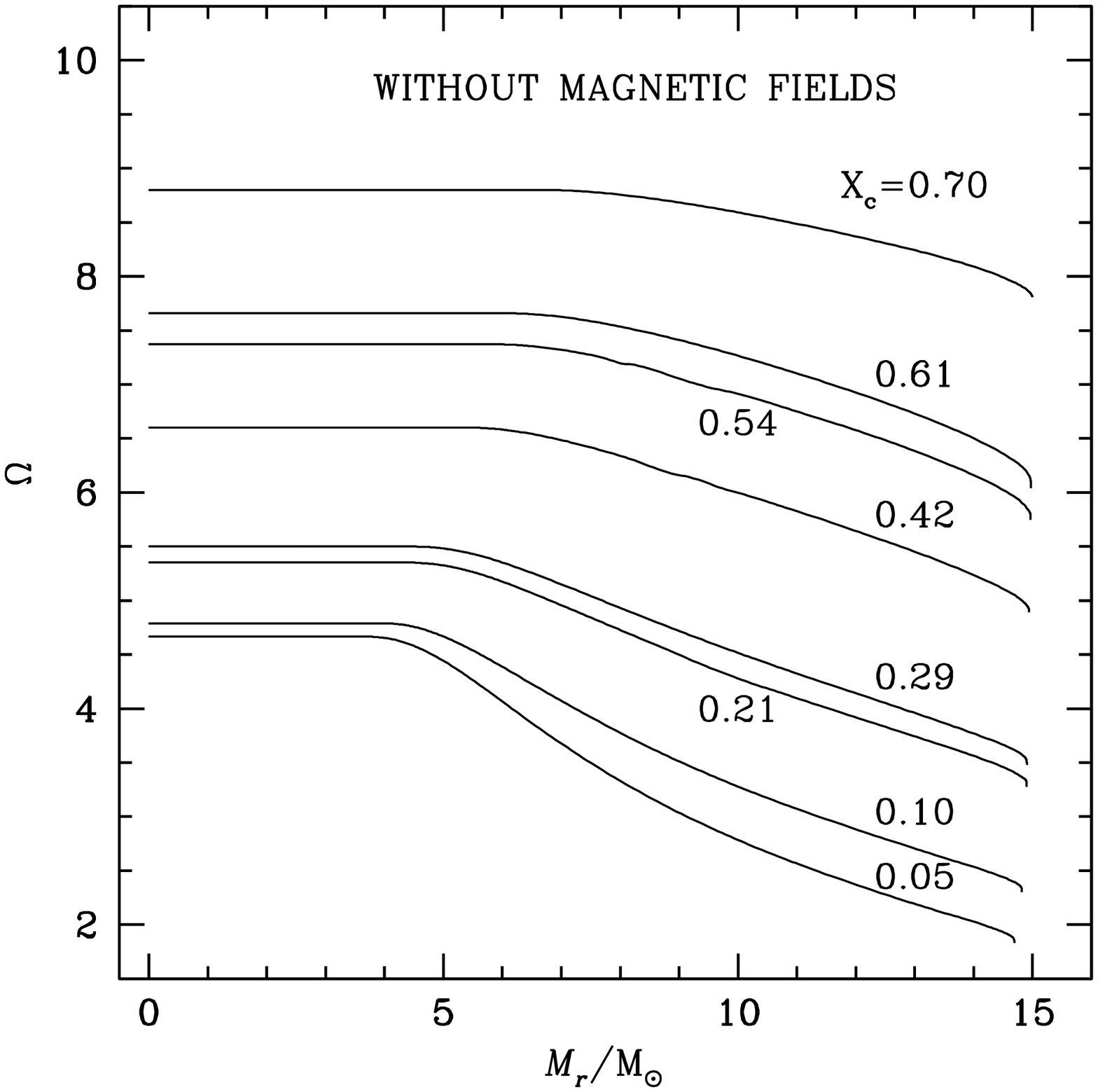}{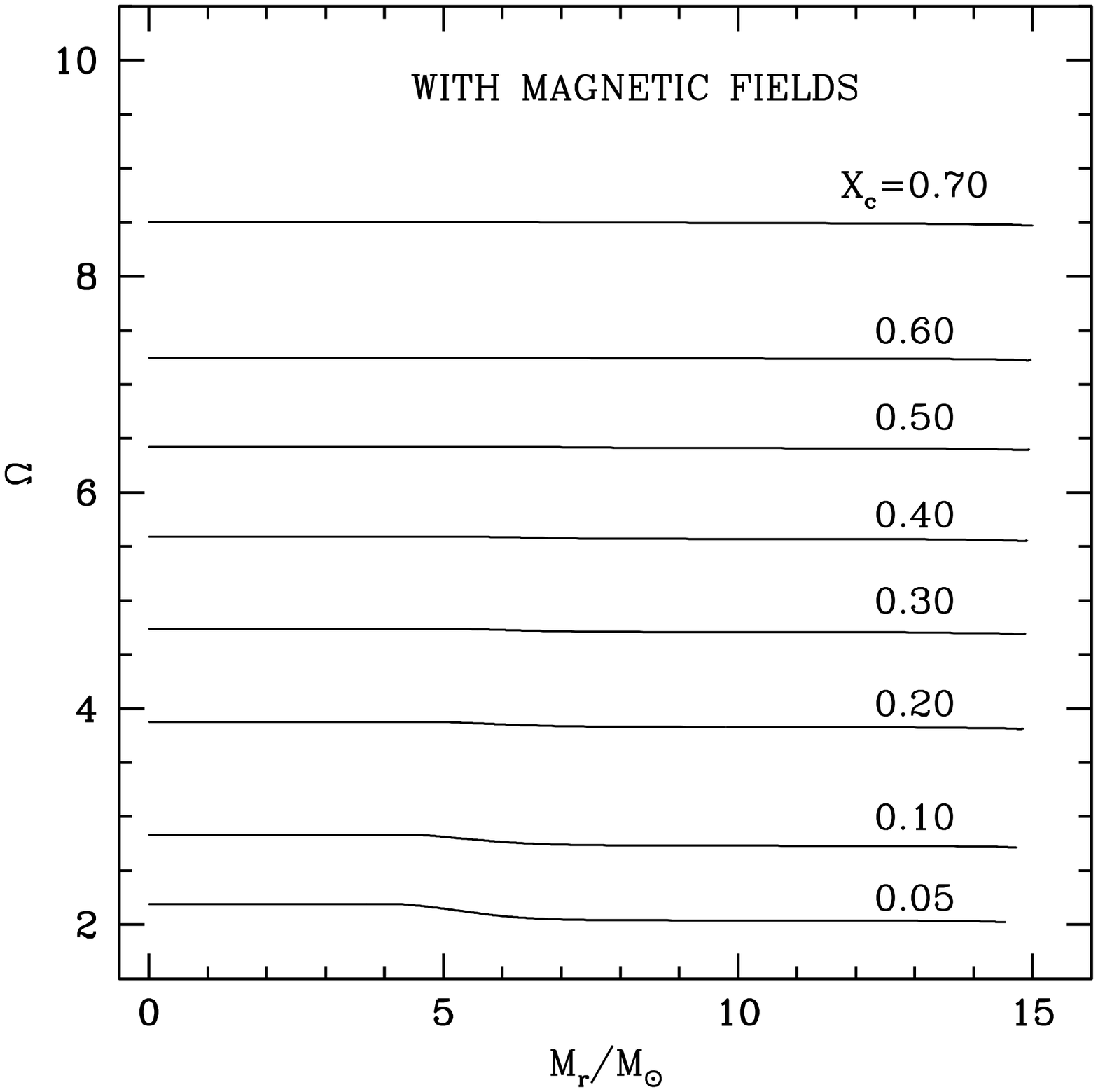}
\caption{{\it Left :} Internal distribution of the angular velocity $\Omega(r)$ 
  as a function of the Lagrangian mass in solar units in a 15 M$_\odot$ model, without magnetic fields, 
  at various stages of the model 
  evolution indicated by the central H--content $X_{\mathrm{c}}$ during the MS--phase. 
  The initial velocity $\upsilon_{\rm ini}$ = 300 km s$^{-1}$.
 {\it Right :} Same as the left figure but with magnetic fields. One notices the almost constant
  values of the angular velocity in the models.}
\label{vh}
\end{figure}

In the interior of stars, at least three mechanisms can transport angular momentum along a radial direction:
\begin{itemize}
\item Convection:here we suppose that convective zones have solid body rotation
\item Meridional circulation:this is the main mechanism
for the transport of the angular momentum in radiative zones.
\item Shear turbulence:only the secular shear turbulence, occurring on thermal timescales, appears
to be important. Dynamical shear, occurring on dynamical
timescales, appears in the advanced stages of the evolution of massive stars and only affects
very locally the profile of $\Omega$ (Hirschi et al. 2003). 
The efficiency of the vertical secular 
shear turbulence for transporting
the angular momentum is in general much smaller than that of the meridional currents. 
\end{itemize}
Let us emphasize that the evolution of the angular velocity inside the star depends on the
gradients of the chemical species and of the angular velocity, these gradients being themselves
deduced from the variation of $\Omega$ as a function of the radius. Thus the problem has to
be solved self-consistently, which is done in the computations presented here.
More precisely, in the models discussed in this section,
the effects of the centrifugal acceleration in the stellar structure equations are accounted for
as explained in Kippenhahn \& Thomas~(1970) (see also Meynet and Maeder~1997).
The equations describing the transport of the chemical species and angular momentum resulting from
meridional circulation and shear turbulence
are
given in Zahn~(1992) and Maeder \& Zahn~(1998) (see  details on the derivation
of the angular momentum transport equation in Meynet \& Maeder 2005b). The expressions for
the diffusion coefficients are taken from Talon \& Zahn~(1997), Maeder~(1997). 
The effects
of rotation on the mass loss rates is taken into account as explained in Maeder \& Meynet~(2000).

\begin{figure}[!t]
\plotfiddle{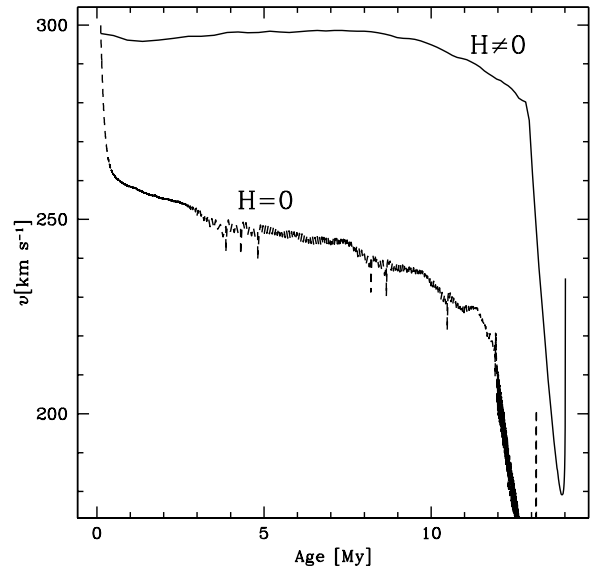}{7cm}{0}{40}{40}{-130}{-60}
\caption{Evolution of the rotation velocities at the surface of 15 M$_\odot$ models 
  during the MS phase with and without magnetic fields. The initial velocity $\upsilon_{\rm ini}$ = 300 km s$^{-1}$ in both cases. 
  We see the much higher surface rotation when magnetic field is included.}
\label{v15}
\end{figure}  

Let us stress also that these models are able to account for many observational constraints that
non--rotating models cannot account for: they can reproduce surface enrichments (Heger \& Langer~2000; Meynet \& Maeder~2000), the blue to red
supergiant ratios at low metallicity \citep{MMVII}, the variation with the metallicity of the Wolf-Rayet populations
and of the number ratios of type Ibc to type II supernovae \citep{MMXI}. 

In the right part of Fig.~\ref{momi}, the evolution of the surface 
velocity for a 20 M$_\odot$ stellar model at solar
metallicity with $\upsilon_{\rm ini}$= 300 km s$^{-1}$ is shown
(see the continuous line). 
Interestingly we note that the evolution of the surface velocity given by consistently taking into
account the above transport mechanisms
is not very far from the solid body rotation case, 
except at the very beginning and 
at the end of the Main Sequence phase. 

At the beginning, the differentially rotating model presents
a decrease of the surface velocity, not shown by 
the solid body rotation model. This initial decrease is due to the action of the
meridional currents, which build up
a gradient of $\Omega$ inside the star, transporting angular momentum from the outer regions to the inner ones.
This slows down the surface of the star.
Then, in the interior
shear turbulence becomes active and
erodes the gradients built by the meridional circulation. Under the influence 
of these two counteracting effects the $\Omega$-profile converges toward an equilibrium configuration. This occurs on a very small timescale
(a few percents of the Main-Sequence lifetime, see Denissenkov et al.~1999). After this
short phase, the variation of $\Omega$ in the radiative zone 
continues to be shaped by shear turbulence, meridional circulation and the change of stellar structure (expansion/contraction of the stellar layers).

At the end of the Main Sequence phase,
when the star is older than about 9 Myr, the surface velocity rapidly decreases. This is
a consequence of the mass loss rate recipe we used in this computation (Vink et al. 2000; 2001) which shows important
enhancement of the mass loss rates when some critical effective temperature are crossed (bistability limits, see the above references). 
In absence of such strong stellar winds, the surface velocity would increase during this phase.

From this computation we can deduce the following results:
first, during the Main-Sequence phase the transport mechanisms are efficient enough to maintain a relatively
weak gradient of $\Omega$ inside the star (see the left part of Fig.~\ref{vh}). The situation is thus not too far from 
the solid body rotation case. Let us however note that the gradients of $\Omega$, although modest, are sufficient enough to drive chemical mixing. These models predict changes of the surface abundances during the Main-Sequence phase well in agreement with
what is observed (Heger and Langer~2000; Meynet \& Maeder~2000; Maeder and Meynet~2001).
Second, this numerical example illustrates the importance of the mass loss in shaping 
the evolution of the surface velocity (see also the discussion below and Fig.~\ref{vz}). 
Third, such models, starting with a higher initial velocity (typically above $\sim$ 350 km s$^{-1}$) would easily reach the critical limit during the
Main-Sequence phase (see the results shown is Meynet and Maeder~2005b).

What does happen after the Main-Sequence phase ? The evolution speeds up and
the variation of $\Omega$ inside the star is mainly governed by the local conservation 
of the angular momentum. We just saw in Sect.~3 above that, in situation where the radius grows up, this makes the 
surface velocity to evolve away from the critical limit. Only when the star contracts, for instance
when a blue loop occurs in the HR diagram, may the star reach the critical limit \citep{HL98}.
This would be a possible scenario for the occurrence of B[e] stars, however
as discussed by Langer \& Heger~(1998), this scenario would predict 
a short B[e] phase (only some 10$^4$ years) with
correspondingly small amounts of mass lost.

Since such stars 
would have had their surface abundances changed by the deep outer convective zone 
appearing at the red supergiant phase, one expects that their surface abundance be 
highly enriched in CNO-processed material: as a numerical example, the N/C and N/O 
ratios at the surface of a 9 M$_\odot$ stellar model at solar metallicity, with an 
initial rotation of $\upsilon_{\rm ini}= 300$ km s$^{-1}$ are enhanced by factors equal to
2.5 and 2.0 respectively during the first crossing of the HR gap from the blue to the red. These two
ratios become 8.5 and 5.3 on the blue loop after a red supergiant stage. The models 
without rotation would predict for these ratios the following values: first crossing, no enhancement for both ratios;
on the blue loop, enhancement factors of 5 and 3.5 respectively \citep{MMX}. This numerical example shows that whatever the initial
rotation velocity, models would predict some CNO-processed material at the surface if the star
comes back from a red supergiant stage, secondly, rotation reinforces the surface enrichment on the blue loop with respect to non--rotating models.

\begin{figure}[!t]
\plottwo{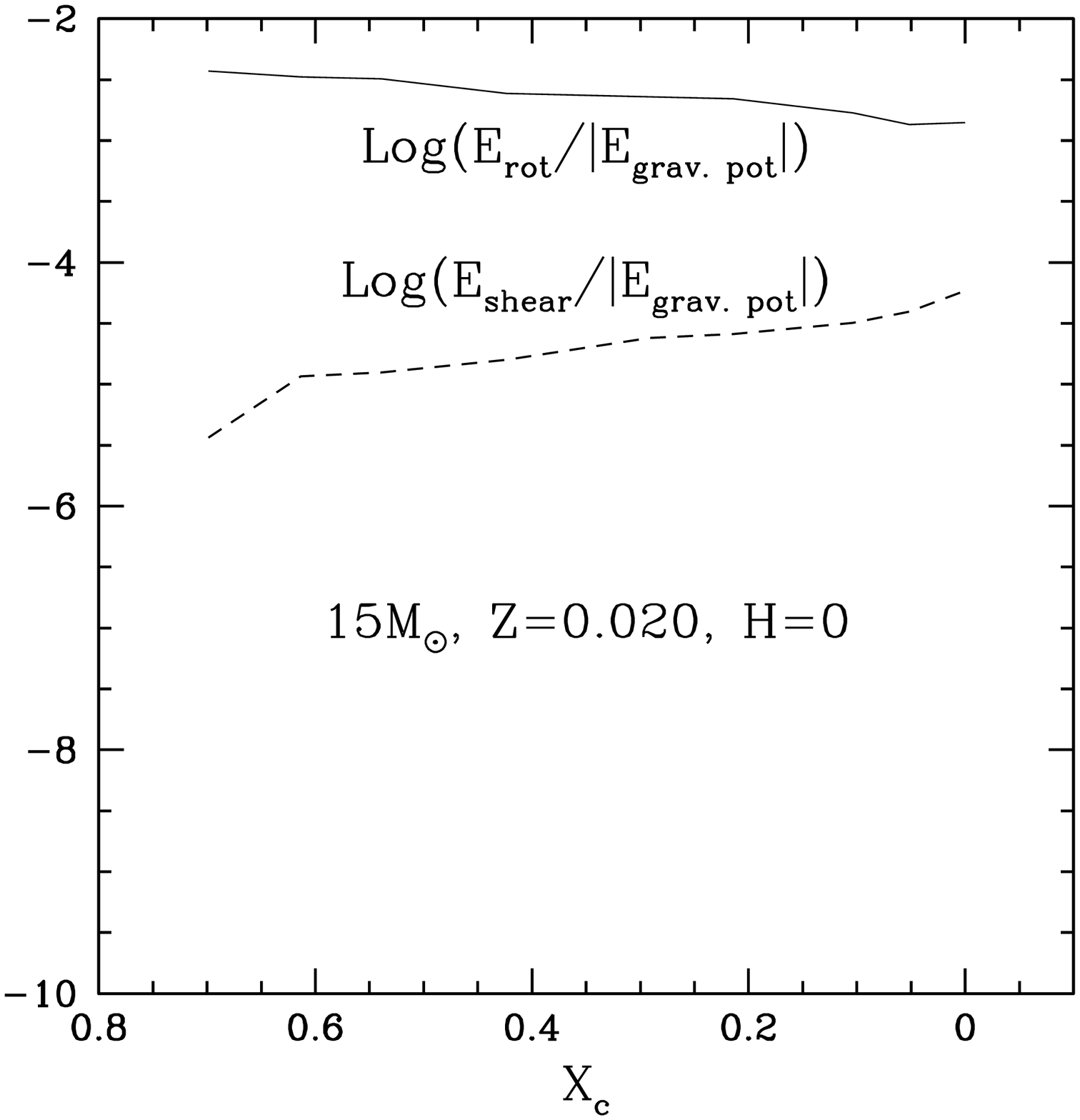}{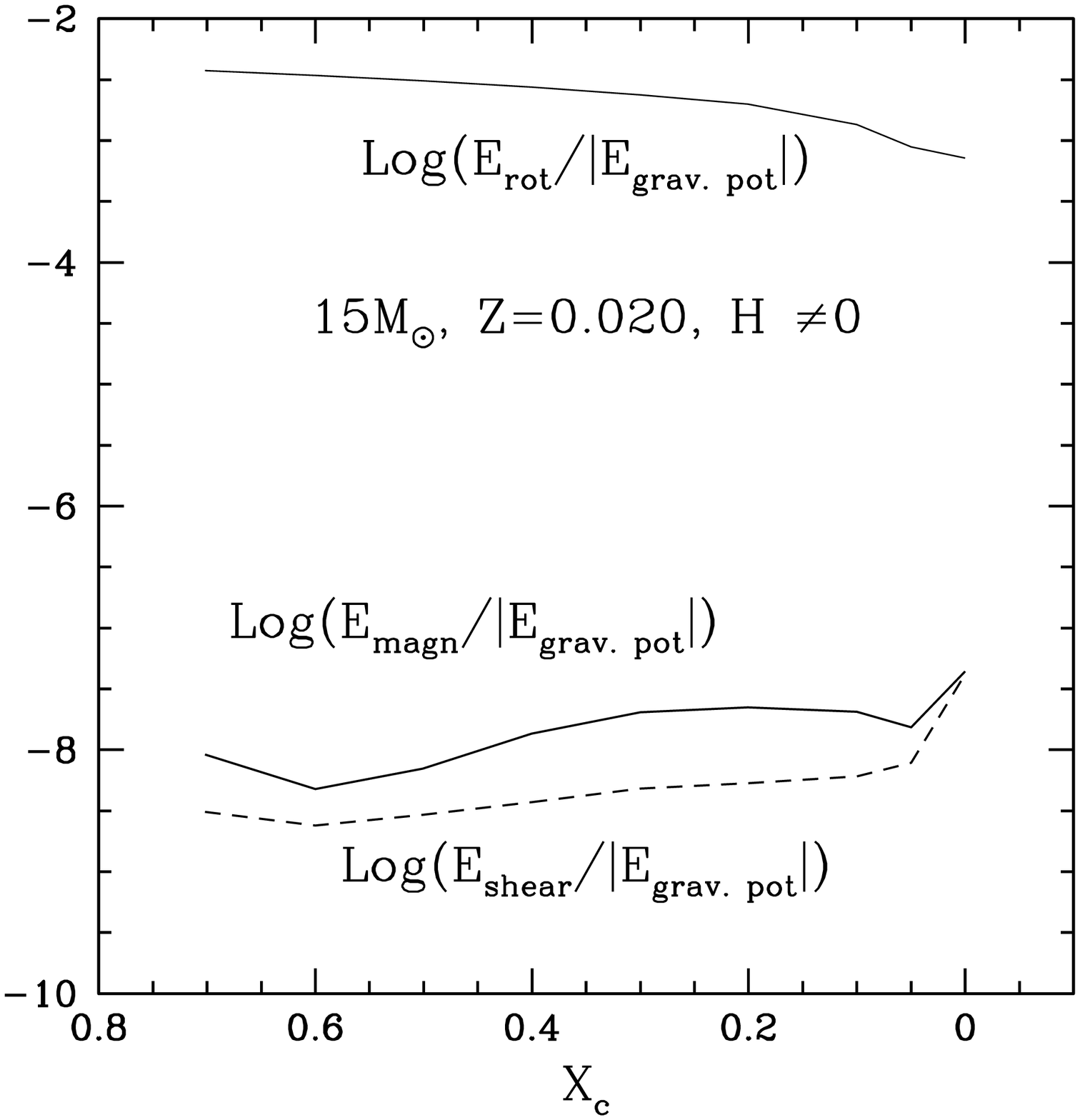}
\caption{ {\it Left :}
Evolution of rotating kinetic energy, of excess 
energy in the shear for a 15 M$_\odot$ stellar model at solar metallicity, 
with $\upsilon_{\rm ini}$ = 300 km s$^{-1}$ without magnetic field. 
The energies are given as fraction of the gravitational energy.
{\it Right :} The same as the left part but for model with rotation and magnetic field. The energy
in the magnetic field is also indicated.}
\label{energie}
\end{figure}

Let us now see what happens to more massive stars. In Fig.~\ref{vz}, the 
evolution of the surface velocity for various 60 M$_\odot$ stellar models at 
two different metallicities is shown. We see that at solar metallicity, the 
mass loss rates are so high that, even starting with an initial velocity of 
500 km s$^{-1}$, the star does not reach the critical limit. The situation is quite different at low $Z$, due
to the metallicity dependence of the mass loss rates (here we used 
$\dot M (Z)=(Z/Z_\odot)^{\alpha}\dot M(Z_\odot)$ with $\alpha$ equal 
to 0.5 as devised by Kudritzki \& Puls~2000). Indeed, at low $Z$ 
little amounts of mass are removed by stellar winds, therefore little amounts 
of angular momentum. As a consequence, the angular momentum brought to the surface 
by the meridional circulation is not removed and accelerate the outer layers. Starting
with $\upsilon_{\rm ini}=$ 300 km s$^{-1}$, the star reaches the critical limit 
at the end of the Main-Sequence phase. Starting with an initial velocity of 800 km s$^{-1}$, the reaching of the critical limit occurs at a much earlier time. 

For still higher initial masses, the stellar luminosity approaches 
the Eddington limit and, as explained above, the critical velocity becomes smaller than the value given by Eq.~(\ref{eqn1}). The star may reach
the $\Omega\Gamma$-limit and lose very high amounts of mass \citep{MMVI}. 
Interestingly, in the HR diagram, the observed position of the de Jager or 
Humphreys-Davidson limit coincides with the position where this $\Omega\Gamma$-limit would occur. 
This may be an indication that in the physics underlying this limit, both rotation and supra-Eddington luminosity play an important role.

\section{Evolution of the surface velocity in models with shellular rotation and magnetic field}

\begin{figure}[!t]
\plotfiddle{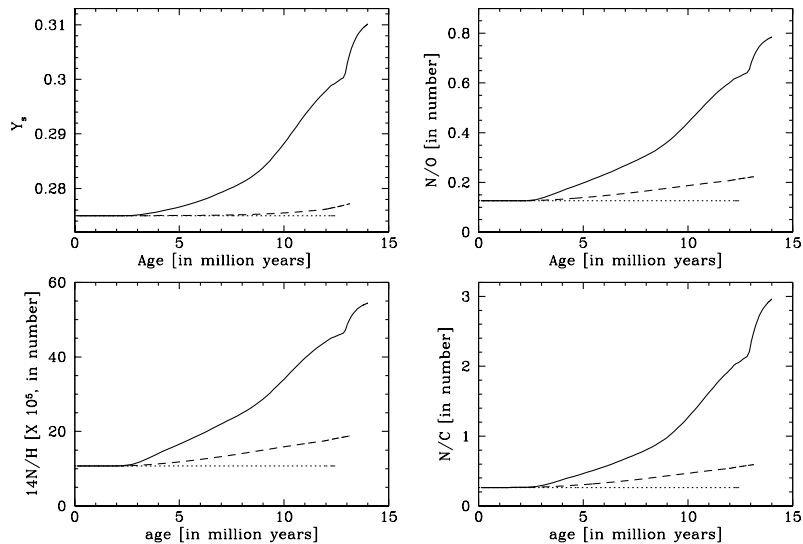}{7cm}{-90}{40}{40}{-160}{250}
\caption{Time evolution of the surface helium content $Y_{\mathrm{s}}$ in mass fraction, of the 
  N/O, N/H and N/C in mass fraction for various models: The dotted line
  applies to  model without rotation, the short--broken line to model with rotation 
  ($\upsilon_{\rm ini}$= 300 km s$^{-1}$) but without
  magnetic fields, the continuous 
  line to model with rotation ($\upsilon_{\rm ini}$= 300 km s$^{-1}$) and magnetic fields. }
\label{abond}
\end{figure}

Spruit~(2002) has proposed a dynamo mechanism operating in 
 stellar radiative layers in differential rotation. This dynamo is based on 
 the Tayler instability, which is the first  one to occur in a radiative zone (Tayler~1973; Pitts \& Tayler~1986). Even a very weak horizontal 
 magnetic field is subject to Tayler instability, which then creates a vertical field component,
 which is wound up by differential rotation. As a result, the field lines become progressively
 closer and denser and thus a strong horizontal field is created at the
 energy expense of differential rotation.
 
 In a first paper \citep{Magn1}, we have shown 
 that in a rotating star a magnetic field can be created during MS evolution 
 by the Spruit dynamo. We have examined the timescale for the field 
  creation, its amplitude and the  related diffusion coefficients. The clear result is that magnetic
  field and its effects are quite important. In the second paper \citep{Magn2}, a generalisation of the  equations of the 
dynamo has been developed. The solutions  fully agree with Spruit's solution in the two limiting
  cases this author has considered \citep{Spruit02}, i.e. 
  ``Case 0'' when the $\mu$--gradient
  dominates and ``Case 1'' when the $T$--gradient dominates with large non--adiabatic effects.
  Our more general solution encompasses all cases of $\mu$-- and $T$--gradients, 
  as well as all cases from the fully adiabatic to  non--adiabatic solutions. In a last paper
  \citep{Magn3}, we examine  the effects of the magnetic field created by Tayler--Spruit dynamo
   in differentially rotating stars. Magnetic fields of the order of a few $10^4$ G are present 
   through most the stellar envelope, with the exception of the outer layers.
   The diffusion coefficient for the transport of angular momentum is 
   very large and it imposes nearly solid body rotation during the MS phase. This can be seen in
   Fig.~\ref{vh} where are compared the evolutions of the angular velocity inside models with and without magnetic fields. 
   
The surface velocities resulting from these two models are shown in Fig.~\ref{v15}. 
Except at the end of the Main Sequence phase, 
the model with magnetic field is strictly equivalent 
to the solid body rotation case. 
   
Does the model with magnetic field predict surface enrichments ? 
Shear turbulence in magnetic models is very weak due to the flatness of the $\Omega$ internal profile. 
The excess of the energy in the shear in the magnetic model is only a few ten thousanths of the 
excess of the energy in the shear in the non--magnetic one (see Fig.~\ref{energie}).
On the other hand,
   solid body rotation
    drives meridional circulation currents which are much faster than usual and leads to much larger 
    diffusion coefficients than the shear diffusivity and than the magnetic diffusivity for the chemical species. As a consequence, 
    the surface enrichments obtained in the models with rotation and magnetic fields are higher than in models with rotation only (see Fig.~\ref{abond}).

\section{Conclusion}

Be stars
might be the natural outcome of stars with initial rotational velocity in the upper tail of the
initial velocity distribution. Depending on when the critical limit is reached, one expects
more or less high surface enrichments. If the critical limit is reached very early during the
Main-Sequence phase, no enrichment is expected, while if the critical limit is reached
at the end of the Main-Sequence phase, high N/C and N/O ratios are expected.
At solar metallicity, for initial masses superior to about 50 M$_\odot$, mass loss rates
prevent the stars to reach the critical limit. The lower initial value for reaching
the critical limit is likely limited by variation of the distribution of the initial velocities.

After the Main-Sequence phase, 
the variation of $\Omega$ inside the star is governed by 
the local conservation of the angular momentum.
In phases during which the radius expands, this makes 
the surface velocity to evolve away from the 
critical limit. In contracting phases, the reverse occurs. 
In the frame of single star evolution models, B[e] could be
in a stage on a blue loop where the star contracted from a previous red supergiant phase.
In that case, the surface is predicted to be enriched in CNO processed material.

The effects of magnetic fields in this context remain to be studied. However, already at this stage,
it appears that magnetic field will facilitate the reaching of the critical limit.






\begin{thebibliography}{}

\bibitem[Denissenkov et al. 1999]{DIW}
Denissenkov, P.A., Ivanova, N.P., Weiss, A. 1991, A\&A, 341, 181

\bibitem[Glatzel(1998)]{Gl98}
Glatzel, W. 1998, A\&A, 339, L5

\bibitem[Heger \& Langer 1998]{HL98}
Heger, A., Langer, N. 1998a, A\&A, 334, 210

\bibitem[Heger \& Langer 2000]{HL00}
Heger, A., Langer, N. 2000, ApJ, 544, 1016

\bibitem[Hirschi et al. 2003]{HMM03}
Hirschi, R., Maeder, A., Meynet, G. 2003, in Stellar Rotation,
IAU Symp. 215, A. Maeder \& P. Eenens (eds.), ASPC, p. 510

\bibitem[Kippenhahn \& Thomas 1970]{KippTh70}
Kippenhahn, R., Thomas, H.C. 1970, in {\it Stellar Rotation}, IAU Coll. 4,
        Ed. A. Slettebak, p. 20
        
\bibitem[Kudritzki \& Puls 2000]{kudpul00}
Kudritzki R.P., Puls J. 2000, ARAA, 38, 613        

\bibitem[Langer 1998]{La98}
Langer, N. 1998, A\&A, 329, 551

\bibitem[Langer \& Heger 1998]{LH98}
Langer, N., Heger, A.  1998, in B[e] stars, A.M. Hubert \& C. Jaschek (eds.), Ap\&SS, 233, 235

\bibitem[Maeder 1997]{Mae97}
Maeder A. 1997, A\&A, 321, 134

\bibitem[Maeder 1999]{Ma99}
Maeder, A. 1999, A\&A, 347, 185

\bibitem[Maeder \& Meynet 2000]{MMVI}
Maeder, A., Meynet, G. 2000, A\&A, 361, 159

\bibitem[Maeder \& Meynet 2001]{MMVII}
Maeder, A., Meynet, G. 2001, A\&A, 373, 555

\bibitem[Maeder \& Meynet 2003]{Magn1} 
Maeder, A., Meynet, G. 2003, A\&A,  411, 543 

\bibitem[Maeder \& Meynet 2004]{Magn2} 
Maeder, A., Meynet, G. 2004, A\&A, 422, 225

\bibitem[Maeder \& Meynet 2005]{Magn3} 
Maeder, A., Meynet, G. 2005, A\&A, 440, 1041

\bibitem[Maeder \& Zahn 1998]{MZ98} 
Maeder A., Zahn J.P. 1998,  A\&A, 334, 1000

\bibitem[Meynet \& Maeder 1997]{MMI}
Meynet, G., Maeder, A. 1997,  A\&A,  321, 465

\bibitem[Meynet \& Maeder 2000]{MMV} 
Meynet, G., Maeder, A. 2000,  A\&A, 361, 101

\bibitem[Meynet \& Maeder 2003]{MMX} 
Meynet, G., Maeder, A. 2003,  A\&A, 404, 975 

\bibitem[Meynet \& Maeder 2005a]{MMXI} 
Meynet, G., Maeder, A. 2005a,  A\&A, 429, 581 

\bibitem[Meynet \& Maeder 2005b]{MMT} 
Meynet, G., Maeder, A. 2005b, in The Nature and Evolution of Disks Around Hot Stars,
R. Ignace and K. G. Gayley (eds.), ASP Conf Ser. 337, p.15 

\bibitem[Pelupessy et al. 2000]{Pe00}
Pelupessy, I., Lamers, H.J.G.L.M., Vink, J.S. 2000, ApJ, 359, 695

\bibitem[Pitts \& Tayler 1986]{Pitts86}
Pitts, E., Tayler, R.J. 1986, MNRAS, 216, 139

\bibitem[Sackmann \& Anand 1979]{Sack70}
Sackmann, I.-J., Anand, S.P.S. 1970, ApJ, 162, 105

\bibitem[Spruit 2002]{Spruit02}
Spruit, H.C. 2002, A\&A, 381, 923

\bibitem[Talon \& Zahn 1997]{TalonZ}
Talon, S., Zahn, J.P. 1997 , A\&A, 317, 749

\bibitem[Tayler 1973]{Tay73}
Tayler, R.J. 1973, MNRAS, 161, 365

\bibitem[Vink et al. 2000]{Vi00}
Vink, J.S., de Koter, A., Lamers, H.J.G.L.M. 2000, A\&A, 362, 295

\bibitem[Vink et al. 2001]{Vi01}
Vink, J.S., de Koter, A., Lamers, H.J.G.L.M. 2001, A\&A, 369, 574

\bibitem[von Zeipel 1924]{vZ24}
von Zeipel, H. 1924, MNRAS, 84, 665

\bibitem[Zahn 1992]{Za92}
Zahn, J.P. 1992, A\&A, 265, 115

\bibitem[Zickgraf 2000]{Zick00}
Zickgraf, F.-J. 2000, in IAU Coll. 175, Myron A. Smith and Huib F. Henrichs (eds), ASPC, 214, p. 26 

\end{thebibliography}
\end{document}